\documentclass[
aps,
prd,
showpacs,
preprint,
tightenlines,
superscriptaddress,
nofootinbib,
floatfix]
{revtex4}
\usepackage{graphicx,
}
\begin{document}
%
\title{     Probing supernova shock waves and neutrino flavor transitions\\
            in next-generation water-Cherenkov detectors }
%
%
\author{        G.L.~Fogli}
\affiliation{   Dipartimento di Fisica
                and Sezione INFN di Bari\\
                Via Amendola 173, 70126 Bari, Italy\\}

\author{        E.~Lisi}
\affiliation{   Dipartimento di Fisica
                and Sezione INFN di Bari\\
                Via Amendola 173, 70126 Bari, Italy\\}

\author{        A.~Mirizzi}
\affiliation{   Dipartimento di Fisica
                and Sezione INFN di Bari\\
                Via Amendola 173, 70126 Bari, Italy\\}
\affiliation	{   Max-Planck-Institut f\"{u}r Physik
				(Werner-Heisenberg-Institut)\\
				F\"{o}hringer Ring 6, 80805 M\"{u}nchen, Germany}

\author{        D.~Montanino}
\affiliation{   Dipartimento di Scienza dei Materiali
                and Sezione INFN di Lecce\\
                Via Arnesano, 73100 Lecce, Italy\\}

\begin{abstract}
Several current projects aim at building a large water-Cherenkov detector,
with a fiducial volume about 20 times larger than in the current Super-Kamiokande 
experiment. These projects include the Underground nucleon decay and Neutrino 
Observatory (UNO) in the Henderson Mine (Colorado), the Hyper-Kamiokande (HK) 
detector in the Tochibora Mine (Japan), and the MEgaton class PHYSics (MEMPHYS)
detector in the Fr{\'e}jus site (Europe). We study the physics potential of 
a reference next-generation detector (0.4 Mton of fiducial mass) in providing 
information on supernova neutrino flavor transitions with unprecedented 
statistics. After discussing the ingredients of our calculations, we compute 
neutrino event rates from inverse beta decay ($\bar\nu_e p\to e^+ n $), 
elastic scattering on electrons, and scattering on oxygen, with emphasis on 
their time spectra, which may encode combined information on neutrino oscillation 
parameters and on supernova forward (and possibly reverse) shock waves. In particular, 
we show that an appropriate ratio of low-to-high energy events can faithfully 
monitor the time evolution of the neutrino crossing probability along the shock-wave 
profile. We also discuss some background issues related to the detection 
of supernova relic neutrinos, with and without the addition of gadolinium.
\end{abstract}
\medskip
 \maketitle

\newcommand{\nuornubar}{{\stackrel{{}_{(-)}}{\nu}\!\!}}
\newcommand{\nubar}{\bar \nu}
\newcommand{\nui}{\nu_i}
\newcommand{\nuj}{\nu_j}
\newcommand{\nubari}{\nubar_i}
\newcommand{\nubarj}{\nubar_j}

\section{Introduction}

The successful operation and the great scientific impact of the Super-Kamiokande 
(SK) water-Cherenkov experiment \cite{Kosh,SKde} have motived several 
research groups to investigate in detail the feasibility of a Megaton-class 
detector \cite{Ko92} for nucleon decay and neutrino physics. Three main projects 
are currently being pursued: the Underground nucleon decay and Neutrino 
Observatory (UNO) project in the Henderson Mine (Colorado, U.S.) \cite{UNNO,Jung}, 
the Hyper-Kamiokande (HK) project in the Tochibora mine (Japan) \cite{HK03}, 
and the MEgaton class PHYSics (MEMPHYS) project in the Fr{\'e}jus site (Europe) 
\cite{Vill}. These next-generation detectors are characterized by a prospective 
fiducial mass of 0.4~Mton or higher,
which will allow studies of neutrinos of astrophysical or terrestrial origin with 
unprecedented statistics and sensitivity \cite{Nu04}.

In particular, a new window would be opened on supernova neutrino physics: 
by naively rescaling the detector mass, it turns out that a 0.4 Mton experiment 
would observe a SN1987A-like signal with a statistics $\sim 200$ times higher
than in Kamiokande (2.14 kton) \cite{Ko92}, and could promote the current 
SK upper limit on supernova relic neutrinos \cite{Male} into a positive detection. 
In general, the observation of a large number of supernova neutrino events will 
allow statistically significant spectral analyses in the energy, angular, and 
time domain. In this context, the identification of relatively model-independent 
spectral features is important to disentangle, as far as possible, information
related to supernova physics and neutrino emission from those related
to neutrino flavor transitions. Recent spectral studies which refer to prospective
Mton-class water-Cherenkov detectors include: analyses in the energy domain to 
probe neutrino mass-mixing parameters \cite{Luna}, to perform multiparameter fits
including neutrino emission parameters \cite{Barg,Mina}, and to identify 
Earth matter effects \cite{Eart}; analyses in the time domain to identify 
the neutronization burst \cite{Whit,Appe} or signatures of shock-wave
propagation effects \cite{Luna,Reve}; analyses in the angular domain to achieve 
supernova pointing \cite{Poin}; background reduction projects using gadolinium 
\cite{GADZ,Ando}.%
\footnote{We mention that large liquid argon detectors (not considered in this work)
can also provide important information on supernova neutrinos; see. e.g., the recent 
works \cite{Arg1,Arg2,Arg3,Arg4,Arg5}. }

In this work, we aim at studying some time dependent and independent observables 
related to (extra)galactic core collapse supernovae, assuming a prospective 
0.4~Mton fiducial mass water-Cherenkov detector. In particular, we investigate 
the effects of the supernova shock propagation on the observable neutrino signal. 
We analyze the imprint of the shock wave on the time spectra of inverse beta decay 
events, and show that the ratio between the number of events in two suitably chosen 
energy ranges can actually monitor the time dependence of the neutrino crossing 
probability $P_H$, thus opening a unique opportunity to study shock-induced flavor
transitions in ``real time''. Moreover, we also study the shock-wave imprint on 
the time spectra of events coming from scattering of neutrinos on oxygen and on 
electrons. We point out that the shock wave produces a characteristic distorsion 
of the oxygen event spectra in time domain. The elastic scattering spectrum is instead  
rather insensitive to the shock-wave propagation and might be thus used to track the 
overall decrease of neutrino luminosity. 

We also study other observables that are basically insensitive to shock-wave effects. 
In particular, we discuss the sensitivity of a 0.4~Mton detector in detecting total event 
rates for various interaction processes, supernova relic neutrinos and neutrinos from 
silicon burning before core collapse.

The structure of this paper is as follows. In Sec.~II we present the ingredients 
of our calculations. We discuss our choice for several input (supernova simulations,
neutrino oscillation parameters, effective detection cross sections) needed 
to compute the SN $\nu$ signal. In  Sec.~III we discuss the results of our calculations 
for observables which are not sensitive to shock waves, i.e., the total number of events 
in the pre-supernova phase and after supernova explosion, and the energy spectra of the 
diffuse SN relic neutrino background with and without gadolinium loading. In Sec.~IV we 
investigate in detail the shock-wave signature on the neutrino time spectra, and present
a possible strategy to monitor the time evolution of the neutrino crossing probability 
along the shock wave profile, by means of an appropriate ratio of low-to-high energy 
events. Conclusions and prospects are given in Sec.~V.

\section{Overview of Calculations}

In this Section we describe the main aspects and ingredients of our calculations 
of supernova neutrino event rates. We remind that, in general, numerical simulations 
of supernova explosions provide the unoscillated double differential neutrino 
distribution in energy and time, 
\begin{equation}
\label{F0nu}
F^0_{\nu}=\frac{d^2 N_\nu}{dE\, dt}\ ,
\end{equation}
where $\nu=\nu_e$, $\overline\nu_e$ and $\nu_x$ in standard notation 
\cite{DiSm,Digh}
($x$ indicating any non-electron flavor).
Such initial distributions are in general modified 
by flavor transitions (see, e.g., \cite{Luna,Digh}),
\begin{equation}
\label{Fnu}
F^0_\nu {\longrightarrow} F_\nu\ ,
\end{equation}
and must be convoluted with the differential interaction cross section 
$\sigma_e$ for electron or positron production, as well as with the detector 
resolution function $R_e$, and the efficiency $\varepsilon$, in order to finally 
get observable event rates \cite{Luna},
\begin{equation}
\label{Conv}
N_e = F_\nu \otimes \sigma_e \otimes R_e \otimes \varepsilon\ .
\end{equation}

\subsection{Neutrino spectra in energy and time}

To our knowledge, only the Lawrence Livermore (LL) group has published in detail \cite{LLLL} 
a successful simulation of supernova explosion and neutrino emission for a 
time interval long enough ($\sim 14$~s) to cover the phenomenon of shock propagation 
\cite{Reve,Schi,DaWi}. Since we are particularly interested to shock-wave signatures 
on neutrino flavor transitions, we assume the LL simulation (which refers to a 
supernova progenitor mass of $\sim 20$ solar masses) as our reference input for 
$t>0$ (time after bounce). 

For $t<0$, Odrzywolek, Misiaszek and Kutschera (OMK) \cite{Sili} have recently estimated 
that the silicon burning phase preceding supernova explosion can release an energy of 
about $5.4\times 10^{50}$~erg in neutrino-antineutrino pairs (with $\nuornubar_e:
\nu_x\simeq 5:1$) for about a couple of days, assuming a 20-solar mass progenitor 
(see, however, \cite{Woos} for different time estimates). In the absence of more detailed 
information, we assume a uniform neutrino flux from silicon burning for two days before 
$t=0$, with emission parameters taken from \cite{Sili}.

\begin{figure}[t]
\vspace*{-0.0cm}\hspace*{-0.2cm}
\includegraphics[scale=0.62]{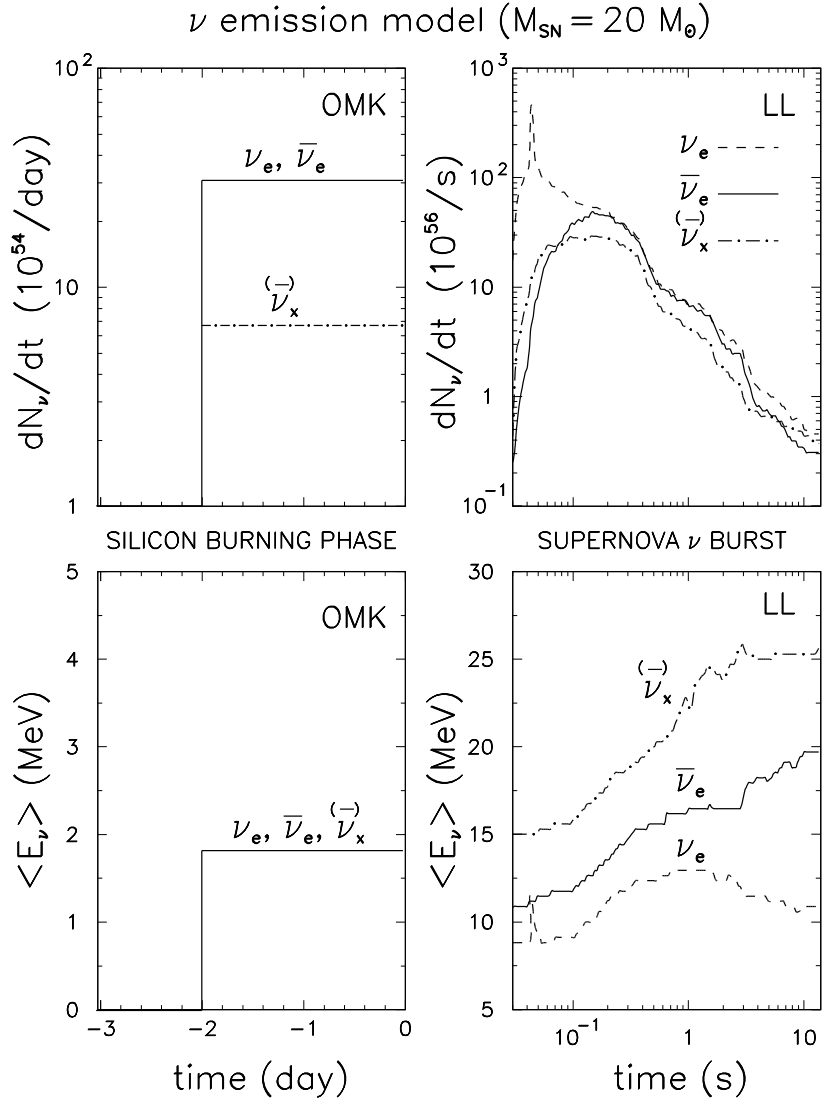}
\vspace*{-0.2cm} \caption{\label{lum}
\footnotesize\baselineskip=4mm Reference neutrino emission parameters as a 
function of time for different flavors: $\nu_e$, $\overline\nu_e$, and $\nu_x$,
assuming a progenitor mass of $\sim 20$ solar masses. Left panels: neutrino emission 
rate $dN_\nu/dt$ and average energy $\langle E\rangle$ for the silicon burning phase, 
as taken from the OMK calculations \cite{Sili}. Right panels: neutrino emission rate 
and average energy for the supernova neutrino burst, as taken from the Lawrence Livermore 
(LL) group simulation \cite{LLLL}.}
\end{figure}

In both cases ($t>0$ and $t<0$) we factorize the differential distribution of 
Eq.~(\ref{F0nu}) as:
\begin{equation}
\label{Fact}
F^0_{\nu}=\frac{dN_\nu}{dt}\,\varphi(E_\nu)\ ,
\end{equation}
for any flavor ($\nu=\nu_e$, $\overline\nu_e$ and $\nu_x$), where $dN_\nu/dt$
represents the neutrino emission rate (number of $\nu$ per unit time), 
while $\varphi(E,t)$ is the
normalized ($\int dE\,\varphi=1$) energy spectrum parametrized as in \cite{Keil}
\begin{equation}
\label{phi}
\varphi(E)=\frac{(\alpha+1)^{\alpha+1}}{\Gamma(\alpha+1)}
\left(\frac{E}{\langle E\rangle} \right)^\alpha
\frac{e^{-(\alpha+1)E/\langle E\rangle}}{\langle E\rangle}\ ,
\end{equation}
where $\langle E\rangle$ is the average neutrino energy and $\alpha$ is an energy
shape parameter. In general, both $\alpha$ and $\langle E\rangle$ can be function 
of time. For simplicity, we have taken $\alpha=3$  for all flavors for $t>0$ 
\cite{Poin}, and $\alpha\simeq4.3$ (from a fit to the spectrum in \cite{Sili}) 
for $t<0$. For $t>0$, the average energies for each flavor are given as a function of
time in the LL simulation \cite{LLLL}, while for $t<0$ they are assumed nearly 
constant and equal to $\sim 1.8$ MeV for all flavors in the OMK evaluation \cite{Sili}. 

Figure~\ref{lum} shows the main characteristics of our reference supernova
emission model. The upper and lower panels show the neutrino emission rate 
$dN_\nu/dt$ and the average neutrino energy $\langle E\rangle$ for different
flavors. The left panels refer to the silicon burning phase from the
OMK calculations \cite{Sili}, while the right panels refer to the supernova neutrino 
burst from the LL group simulation \cite{LLLL}. According to the LL simulation, all 
flavors have comparable emission rates (within a factor $\sim 2$) in the so-called 
cooling phase ($t\gtrsim 0.5~s$). In the preceding phase the relative 
$\nu_e$ emission rate is higher, and shows a distinct neutronization peak at 
0.04--0.05 s. For later purposes, we observe that the LL simulations predict 
limited variations ($<20\%$) of the average neutrino energy in the time range 
($t>2$~s) relevant for shock-wave effects on neutrino flavor transitions.

\begin{figure}[t]
\vspace*{-0.0cm}\hspace*{-0.2cm}
\includegraphics[scale=0.75]{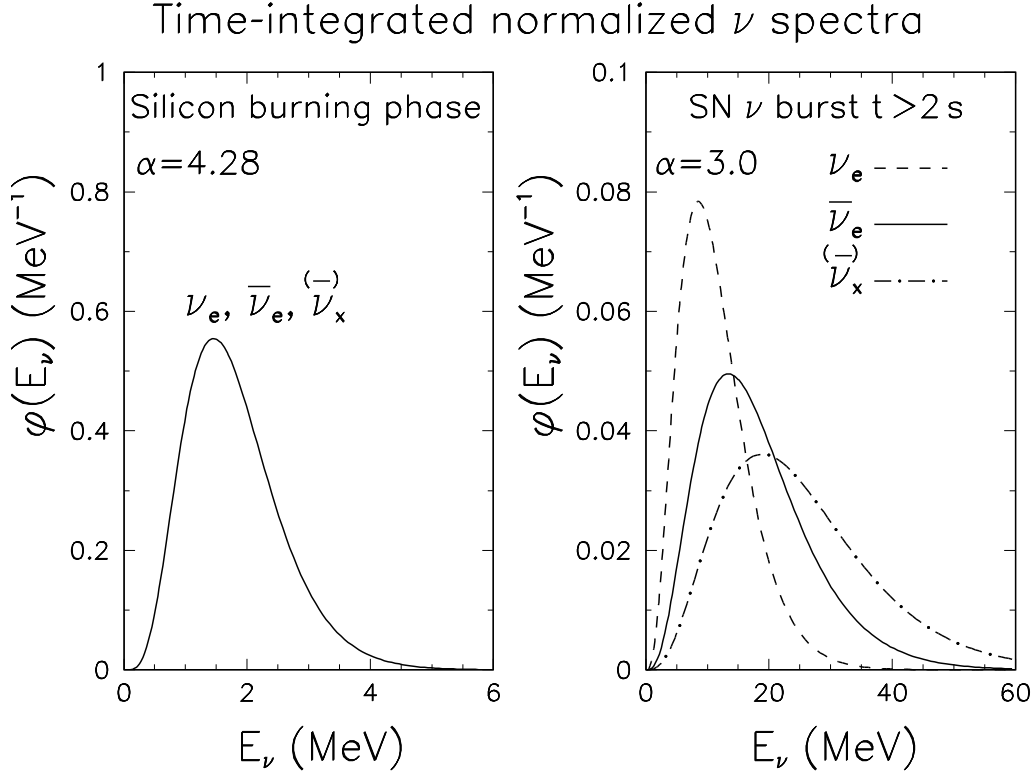}
\vspace*{-0.2cm} \caption{\label{spe}
\footnotesize\baselineskip=4mm Normalized neutrino spectra averaged over time. 
Left panel: energy spectra in the silicon burning phase, as derived from 
\cite{Sili} with shape parameter $\alpha=4.28$ \cite{Keil}. Right panel: energy 
spectra associated to the supernova neutrino burst, averaged in the time interval 
for shock-wave effects ($t>2$), as derived from \cite{LLLL} with shape parameter 
$\alpha=3$ \cite{Keil}.}
\end{figure}

Figure~\ref{spe} shows the normalized neutrino spectra $\varphi$ for each flavor,
after a flux-weighted average over time, for both the silicon burning phase (left panel) 
and after supernova explosion (right panel). For later purposes, in the right panel we 
have restricted the integration interval to $t>2$~s, i.e., to the interval where shock-wave 
effects are relevant for flavor transitions.  Note that, in the silicon burning phase, 
the spectra are equal for all flavors, and are peaked at relatively low energy ($<2$~MeV) 
\cite{Sili}. During the supernova explosion, neutrinos have an order of magnitude higher 
average energy, and are peaked at different energy for each flavor \cite{LLLL}. 
Of course, the detailed features of the reference emission model in Figs.~\ref{lum} 
and~\ref{spe} must be taken with a grain of salt, since the distribution of the total
energy in flavor and time is currently subject to large uncertainties, which may be 
reduced in more advanced future simulations  \cite{Bura}.

\subsection{Neutrino crossing probability and shock waves}

Assuming standard three-neutrino mixing, the parameters relevant to flavor
transitions in supernovae are the two independent squared mass difference $\delta m^2$
and $\Delta m^2$ and the mixing angles $\theta_{12}$ and $\theta_{13}$ (see, 
e.g., \cite{Simp}). For $\delta m^2\simeq 8\times 10^{-5}$ eV$^2$, as currently
indicated by reactor and solar neutrino data \cite{KamL}, the dependence on $\delta m^2$
actually vanishes, while the dependence on $\Delta m^2$ and on $\theta_{13}$ can be 
essentially embedded in the so-called crossing probability $P_H$ (up to Earth matter 
effects that, for simplicity, we do not consider in this work). In general, $P_H$ 
takes the same form for both neutrinos and antineutrinos \cite{Simp},
\begin{equation}
\label{PH}
P_H=P_H(\Delta m^2/E,\sin^2\theta_{13},V(x,t))\ ,
\end{equation}
where $V$ is the neutrino potential profile \cite{Matt} at radius $x$ and time $t$, 
\begin{equation}
\label{V}
V(x,t)=\sqrt{2}\,G_F\,N_e(x,t)\ ,
\end{equation}
and $N_e$ is the electron density. When needed, we fix 
$\Delta m^2\simeq 2.4\times 10^{-3}$~eV$^2$ 
from atmospheric and accelerator data \cite{SKam}, $\sin^2\theta_{12}\simeq 0.3$ 
from reactor and solar data \cite{KamL}, and take representative value of 
$\sin^2\theta_{13}$ below current upper bounds \cite{CHOO}.

Matter effects are potentially relevant when the potential equals the neutrino 
wavenumber $k_H=\Delta m^2/2E$,
\begin{equation}
\label{Vk}
V(x,t)\simeq k_H\ .
\end{equation}
Indeed, it has been shown in \cite{Miri} that, even for non-monotonic supernova density 
profiles, the evaluation of $P_H$ in Eq.~(\ref{PH}) can be obtained with good approximation 
through a simple procedure involving the ordered product of matrices embedding the local 
crossing probabilities \cite{Petc}, evaluated at all points where Eq.~(\ref{Vk}) is fulfilled. 
This approximation is particularly useful to study the effect of non trivial density profiles,
such as those induced by shock waves.

In a seminal paper \cite{Schi}, Schirato and Fuller noted that the propagation of the forward 
shock wave in the LL simulation \cite{DaWi} could influence neutrino flavor transitions a 
few seconds after core bounce. 
In addition to forward shock effects, it has been recently pointed out 
\cite{Reve} that a second (``reverse'') shock front, is also expected to propagate behind the 
forward one (at a lower velocity), although a detailed description is not yet possible 
within current numerical experiments. 

\begin{figure}[t]
\vspace*{-0.0cm}\hspace*{-0.2cm}
\includegraphics[scale=0.55]{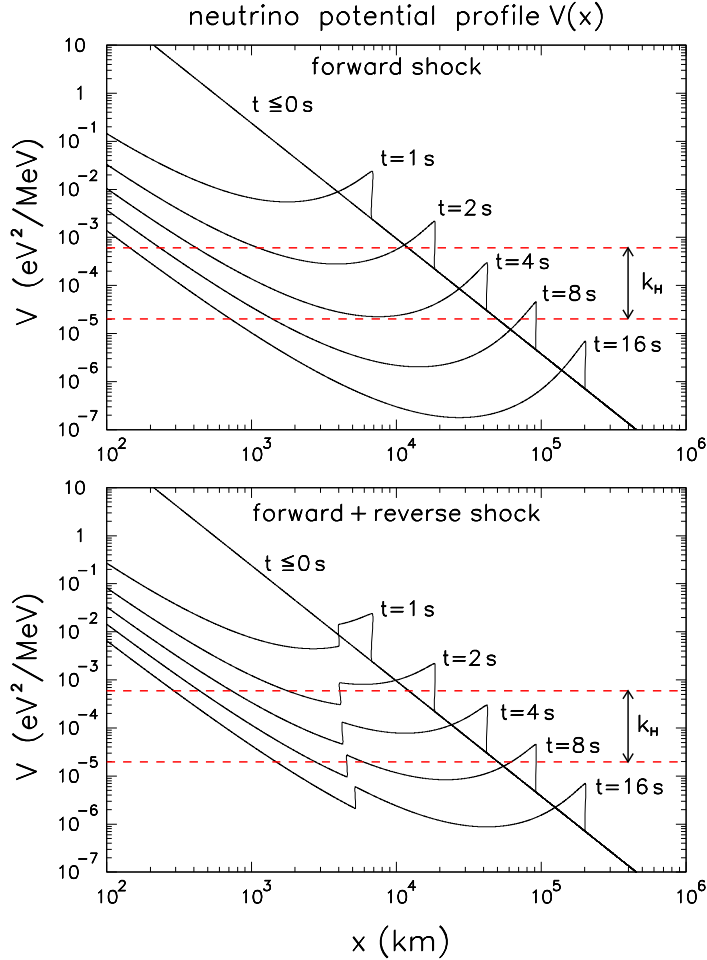}
\vspace*{+0.2cm} \caption{\label{pot}
\footnotesize\baselineskip=4mm Radial profile of the neutrino potential $V(x)$ at different 
times $t$ (1, 2, 4, 8, and 16 s). Upper panel: Our simplified profile for the case of forward shock
\cite{Schi} (see also \cite{Miri}). Lower panel: Our simplified profile for the case of forward plus 
reverse shock, adapted from \cite{Reve}. In both cases, the static profile ($t\leq 0$~s) is
also shown. The band within dashed lines marks the region where matter effects are potentially 
important ($V\simeq k_H$ for $E=2$--60~MeV).}
\end{figure}

Figure~\ref{pot} shows the simplified shock-wave profiles used in this work. 
The upper panel refers to the neutrino potential $V$ in the presence of forward
shock only \cite{Schi}, using the same parametrization as in \cite{Miri}. 
The main features of the forward shock profile are a sharp discontinuity at the shock 
front (which can induce a strongly nonadiabatic transition) leaving behind
an extended rarefaction zone. In the lower panel we have graphically 
adapted the results of the simulation in \cite{Reve} to account for a reverse shock,
characterized by a smaller discontinuity at the front. In both panels, 
we also show the band spanned by the neutrino wavenumber $k_H=\Delta m^2/2E$
for $E\in[2,60]$ MeV. The condition for large matter effects ($V\simeq k_H$)
implies then that at relatively early (late) times only the static profile
(the rarefaction zone) is relevant, while for intermediate times (e.g., 
$t=$2, 4 or 8~s in Fig.~\ref{pot}) the propagation of the shock front(s) must
be accounted for.%
\footnote{Neutrinos from silicon burning may also experience 
flavor transitions along the static profile (not considered in \cite{Sili}). 
In fact, for a typical energy $E\simeq2$~MeV
(see Fig.~\ref{spe}, left panel),
the condition $V(x)\simeq k_H$ is fulfilled at $x\simeq 10^4$~km for the static profile,
i.e., below the estimated silicon core radius ($x_\mathrm{Si}\sim 
4\times 10^{3}$~km).}

\begin{figure}[t]
\vspace*{-0.0cm}\hspace*{-0.2cm}
\includegraphics[scale=0.55]{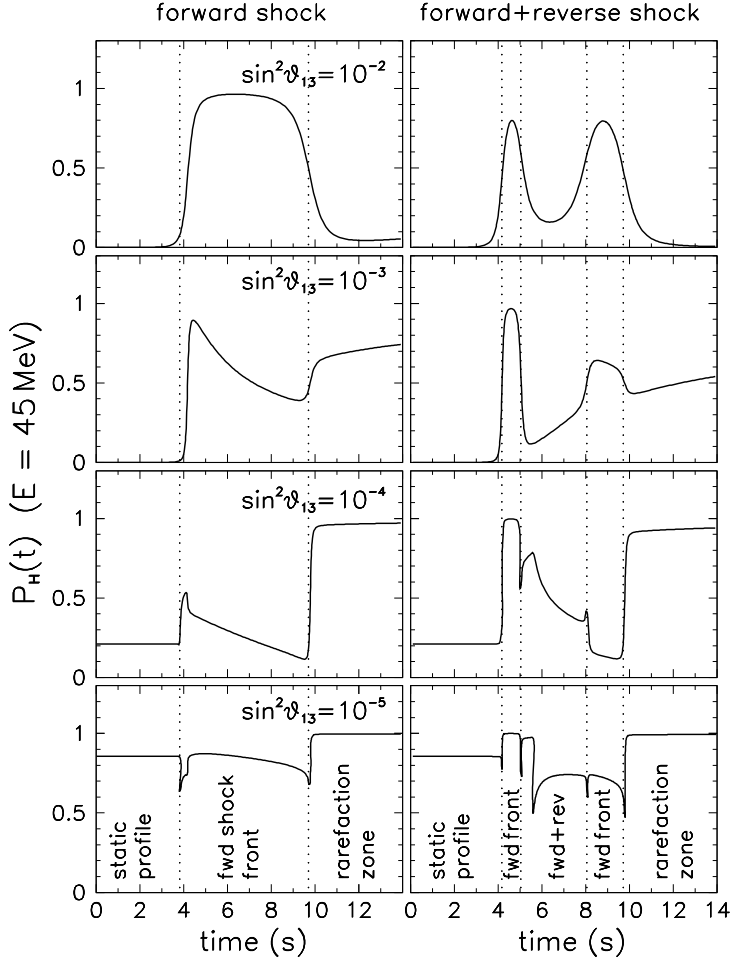}
\vspace*{-0.2cm} \caption{\label{pro}
\footnotesize\baselineskip=4mm Crossing probability $P_H$ as a function of time,
at relatively high neutrino energy ($E=45$~MeV). Left panels: case with
forward shock only. Right panels: case with forward plus reverse shock. From top
to bottom, the mixing parameter $\sin^2\theta_{13}$ takes the values $10^{-2}$,
$10^{-3}$, $10^{-4}$, and $10^{-5}$. The function $P_H(t)$ changes rapidly 
at the times indicated by dotted vertical lines, i.e., when the 
static profile is first perturbed by the forward shock front (followed by
the reverse shock front in the right panels) and then by the rarefaction zone.
Notice how the reverse shock partially ``undoes'' the $P_H$ variation
induced by the forward shock.}
\end{figure}

Figure~\ref{pro} shows our calculation of $P_H$ as a function of time for four 
representative values of $\sin^2\theta_{13}$, and for a relatively high neutrino energy 
($E=45$~MeV, useful for later purposes). The left panels refers to the case with forward 
shock only, where the strong variations induced by the passage of the front discontinuity 
and then by the rarefaction zone are marked by vertical dotted lines. We refer the reader 
to \cite{Miri} for a thorough discussion of forward-shock effects on $P_H$. The right panels 
refer to the case with forward plus reverse shock, where the effects of the reverse shock 
appear to change dramatically the crossing probability at intermediate times. Qualitatively, 
the nonadiabatic transition at the reverse shock front can partially ``undo'' the effect 
of the analogous transition at the forward shock front; this is particularly evident, e.g., 
in the two upper panels. We refer the reader to \cite{Reve} for further discussions about the 
time dependence of $P_H$ in the presence of a reverse shock. Despite its complexity, such 
time dependence might be monitored surprisingly well in future water-Cherenkov detectors,
as we shall see in Sec.~IV.

\subsection{Neutrino spectra and ``critical energy''}

The neutrino fluxes at the supernova exit ($F_\nu$) are linear combinations of the 
initial fluxes ($F_\nu^0$), with coefficients governed by $P_H$ and by the 
mixing angle $\theta_{12}$, as well as by the mass spectrum hierarchy (normal or inverted). 
In particular, for normal mass hierarchy it is (see, e.g., \cite{Digh})
\begin{eqnarray}
F_{\overline\nu_e} &\simeq& \cos^2\theta_{12} F^0_{\overline\nu_e}+
\sin^2\theta_{12}F^0_{\nu_x}\ , 
\label{FnubarNH}\\
F_{\nu_e} &\simeq & \sin^2\theta_{12} P_H F^0_{\nu_e}+
(1-\sin^2\theta_{12}P_H)F^0_{\nu_x}
\label{FnuNH}\ , 
\end{eqnarray}
while, for inverted hierarchy,
\begin{eqnarray}
F_{\overline\nu_e} &\simeq & \cos^2\theta_{12} P_H F^0_{\overline\nu_e}+
(1-\cos^2\theta_{12} P_H)F^0_{\nu_x}\ , 
\label{FnubarIH}\\
F_{\nu_e} &\simeq& \sin^2\theta_{12} F^0_{\nu_e}+
\cos^2\theta_{12}F^0_{\nu_x}
\label{FnuIH}\ .
\end{eqnarray}
Notice that, for $\overline\nu_e$, the case of inverted hierarchy with $P_H=1$ is 
indistinguishable from the case of normal hierarchy; similarly, for $\nu_e$, the case 
of normal hierarchy with $P_H=1$ is indistinguishable from the case of inverted hierarchy. 

For later purposes, it is useful to introduce the concept of ``critical energy'' $E_c$ 
\cite{Luna}, defined as the energy where the initial $\overline\nu_e$ and $\nu_x$ fluxes 
are approximately equal,
\begin{equation}
\label{Ec}
F^0_{\overline\nu_e}(E_c)\simeq F^0_{\nu_x}(E_c)\ .
\end{equation}
At the critical energy, flavor transitions are not effective in the antineutrino
channel, and the $\overline\nu_e$ flux at supernova exit equals the 
initial one,
\begin{equation}
\label{Ec2}
F_{\overline\nu_e}(E_c)\simeq F^0_{\overline\nu_e}(E_c)
\end{equation}
for both normal and inverted hierarchy.

In the LL simulation \cite{LLLL}, the critical energy is relatively well defined 
and stable for $t>2$~s (the range relevant for shock-wave effects); in this time 
interval, the $\overline\nu_e$ and $\nu_x$ emission rates are comparable, and their 
average energies do not vary too much in time (see Fig.~\ref{lum}). Within the LL 
simulation, the critical energy is then approximately defined by the crossing point
of the time-averaged $\overline\nu_e$ and $\nu_x$ spectra in Fig.~\ref{spe}, i.e., 
$E_c\sim 20$~MeV. This energy provides a sort of ``no oscillation benchmark'', in 
comparison with higher energies (say, $E\sim 40$--50~MeV), where $F^0_{\nu_x}\gg 
F^0_{\overline\nu_e}$ (see Fig.~\ref{spe}) and flavor oscillation effects do not 
cancel. Of course, the next real supernova explosion might be characterized by a 
``critical energy'' different from this or other simulations.%
\footnote{A remark is in order. The concept of ``critical energy'' is well defined
only if the $\bar\nu_e$ and $\nu_x$ energy spectra turn out to have 
different
shapes and thus a crossing energy point, as is the case for the LL
simulation \cite{LLLL} at $E_c\simeq 20$~MeV.
Recent simulations \cite{Bura} predict $\bar\nu_e$ and $\nu_x$ spectrum
differences which,  although smaller than in the LL case, in general still allow
a meaningful definition of the critical energy.}
 We shall discuss in 
Sec.~IV~A a possible way to circumvent our ``a priori'' ignorance of the real value 
of $E_c$, when seeking signatures of shock-wave effects.

\subsection{Neutrino interactions and detection}

In large water-Cherenkov detectors, the interaction processes which can provide observable 
supernova neutrino event rates are: (i) inverse beta decay (possibly with neutron capture 
in Gd \cite{GADZ}); (ii) (anti)neutrino scattering on oxygen; and (iii) 
(anti)neutrino scattering 
on electrons. For the process (i), we take the differential cross section from \cite{Stru}
(we recall that the positron energy $E_\mathrm{pos}$ closely tracks the neutrino
energy $E$ \cite{Stru}). For the process (ii), we take the total 
cross section from \cite{Kolb}, where we have assumed the same energy distribution for all the 
channels. The differential cross sections for the processes in (iii) are well-known (see, e.g., 
\cite{Poin}).

For any interaction process, we fold the differential cross sections for $e^{\pm}$ production 
with a Gaussian energy resolution function of width $\Delta$, and apply a sharp cut to events 
with measured $e^\pm$ energy below a threshold value $E_\mathrm{thresh}$. The value of $\Delta$ 
is determined by the photocatode coverage. For a coverage comparable to the current one in 
Super-Kamiokande~II ($\sim 20\%$ of the area \cite{Suzu}), as envisaged in typical
Mton-class projects \cite{Nu04}, we derive from the information in \cite{Suzu} that
\begin{equation}
\label{Delta}
\Delta/\mathrm{MeV} \simeq 0.6\sqrt{E/\mathrm{MeV}}\ .
\end{equation}
We also adopt the conservative value $E_\mathrm{thresh}=7$~MeV from \cite{Vill}. The detection 
efficiency $\varepsilon$ is assumed to be 1 above threshold. Unless otherwise noticed, the  
fiducial volume for supernova neutrino detection is assumed to be 0.4 Mton \cite{UNNO,Vill}.

\begin{figure}[t]
\vspace*{-0.0cm}\hspace*{-0.2cm}
\includegraphics[scale=0.74]{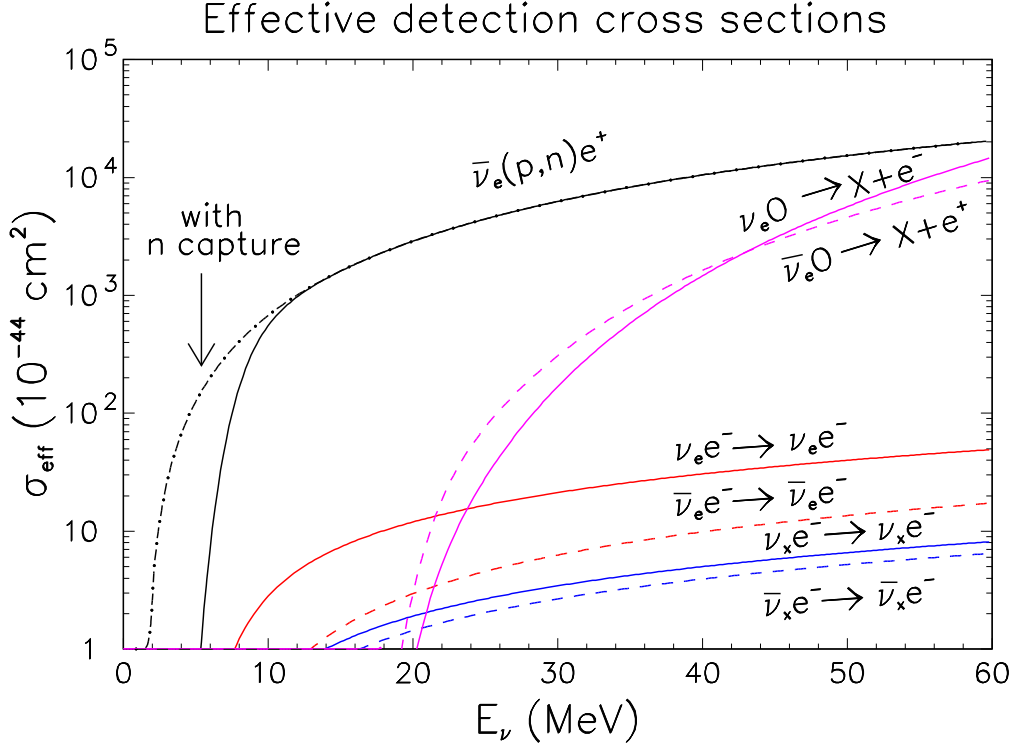}
\vspace*{-0.2cm} \caption{\label{xse}
\footnotesize\baselineskip=4mm Effective neutrino interaction cross sections
as a function of energy, including energy resolution and threshold effects. See
the text for details.}
\end{figure}

Figure~\ref{xse} shows the effective cross sections as a function of neutrino energy for the
various processes discussed before, including resolution and threshold effects. Notice that, by 
adding Gd \cite{GADZ}, the neutron capture signature allows to detect sub-threshold inverse beta 
decay events.

\section{Observables not probing shock waves}

In this Section we discuss the results of our calculations for time-integrated observables, such 
as the total number of events from a single supernova explosion or from the diffuse supernova 
neutrino background, which are basically insensitive to shock-wave effects. Oscillation effects 
are simply accounted for by varying $P_H$ in its full range $[0,1]$ for both normal and inverted 
hierarchy (for a static profile, this is equivalent to vary $\sin^2\theta_{13}$ in the currently 
allowed range, $\sin^2\theta_{13}\lesssim$~few\%, see e.g., \cite{Luna}).

\begin{figure}[t]
\vspace*{-0.0cm}\hspace*{-0.2cm}
\includegraphics[scale=0.88]{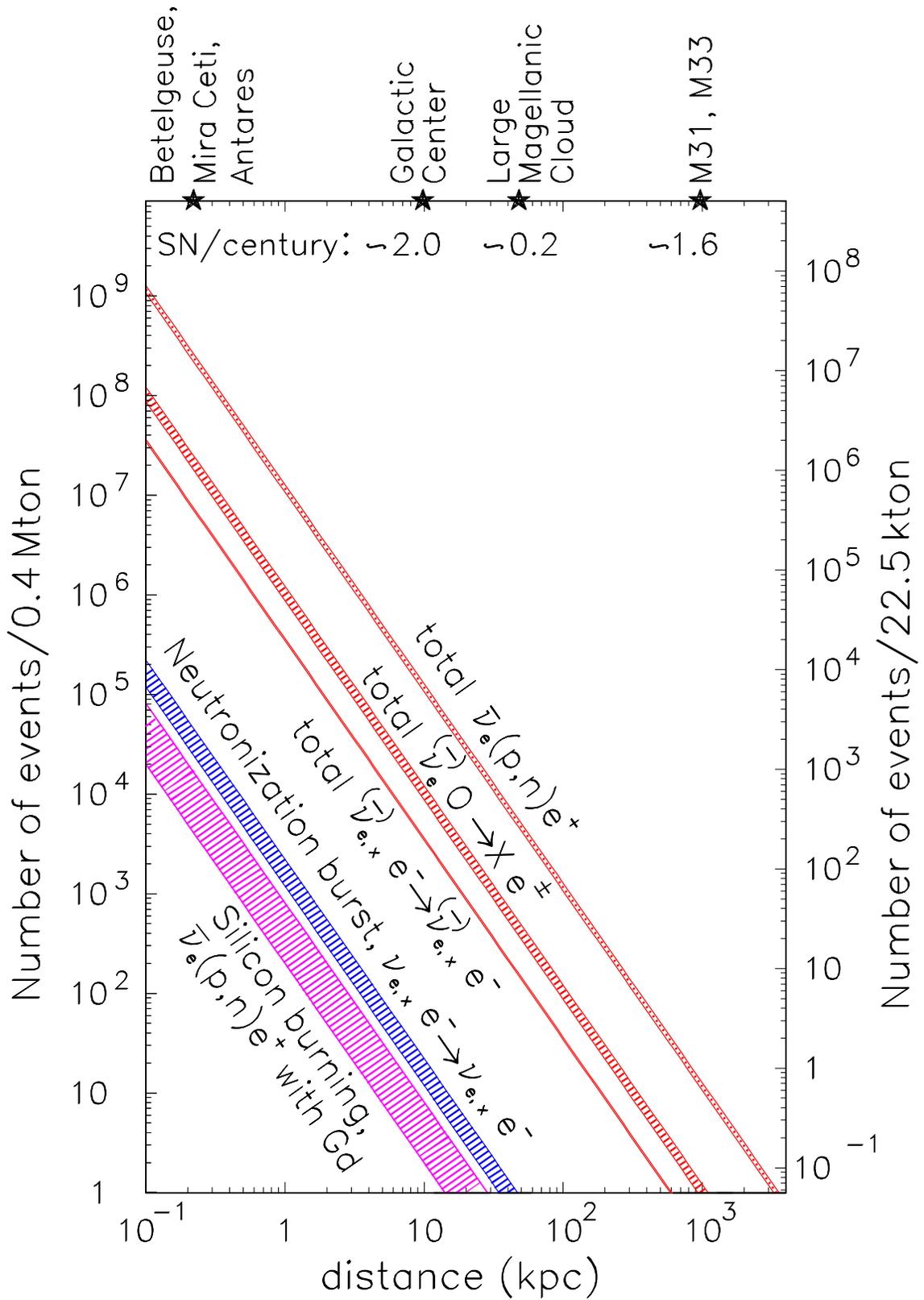}
\vspace*{-0.2cm} \caption{\label{log}
\footnotesize\baselineskip=4mm Number of events expected in 0.4 Mton detector (left y-axis) or in a 
SK-like detector (right y-axis) as a function of the supernova distance, for various interaction 
channels. See the text for details.}
\end{figure}

\subsection{Total number of events}

Figure~\ref{log} shows the total number of events expected in a 0.4 Mton detector for various 
interaction processes, as a function of the supernova distance in kpc. (For the sake of comparison, 
the vertical axis on the right refers to the number of events for a SK-like volume of 22.5 kton). 
Relatively close stars which might evolve into core-collapse supernovae at unpredictable future 
times include Betelgeuse, Mira Ceti, and Antares. For close galaxies \cite{Tull}, the rate of 
core-collapse supernovae may be inferred from their luminosity and morphology \cite{Capp}; 
representative estimates based on \cite{Tull,Capp} (to be taken within a factor of two) are reported
in the upper part of the figure for: the galactic center, the Large Magellanic Cloud (LMC), and the 
M31+M33 spiral galaxies. For each interaction process, the uncertainty due to our ignorance of 
the hierarchy and of $P_H\in[0,1]$ is accounted for by the vertical spread of the bands. The 
relatively different spreads reflect different sensitivities to neutrino flavor transitions. 
Let us now discuss the various classes of events from bottom to top.

Inverse beta decay events from silicon burning have very low (sub-threshold) positron energies, 
and can only be detected through neutron capture by adding gadolinium \cite{GADZ}, provided 
that they can be statistically distinguished from background fluctuations. At a distance of 1~kpc, 
we estimate $\sim 200$ to $\sim 800$ Si burning events over about two days. This large variation is 
due to the large ratio $F^0_{\overline\nu_e}:F^0_{\nu_x}\simeq 5:1$ (see Sec IA), which implies 
$F_{\overline\nu_e}\sim 0.2$--$0.8\times F^0_{\overline\nu_e}$, where the lowest value is obtained 
for inverted hierarchy and $P_H=0$, while the highest value is obtained for either inverted hierarchy 
with $P_H=1$ or for normal hierarchy. The estimated signal rate of $\sim 100$--$400$ events/day at 
1~kpc should be compared with the background rate from spallation neutrinos (e.g., 1780 events/day 
at the Kamioka site), reactor neutrinos (e.g., $\sim 500$ events/day at the Kamioka site), and from 
minor contributions (natural radioactivity). We have taken such backgrounds from 
\cite{GADZ} and rescaled them to a 0.4~Mton mass. Assuming a typical total background rate of 
$\sim 2500\pm 50$ events per day (statistical errors only), the silicon burning signal should then 
be seen with a statistical significance of 2--8 standard deviations at a reference distance of 1 kpc. 
Unfortunately, at the galactic center ($\sim 10$ kpc) the estimated silicon burning signal would be 
100 times smaller and thus unobservable. 

There are better prospects to observe the neutronization burst from a galactic supernova by means 
of elastic scattering on electrons, including contributions from all flavors. The contribution of 
antineutrinos can be neglected, being suppressed both by the original flux and by the cross section 
(see Figs.~\ref{lum} and \ref{xse}). The contribution of non-electron neutrino flavors can be derived 
by using Eq.~(\ref{FnuNH}) or Eq.~(\ref{FnuIH}) plus the unitarity condition $F_{\nu_e}+F_{\nu_\mu}+
F_{\nu_\tau}=F^0_{\nu_e}+2F^0_{\nu_x}$. In general, the dominant contribution comes from $F_{\nu_e}$ 
(with $F_{\nu_e}^{\mathrm{max}}\simeq F^0_{\nu_e}\sin^2 \theta_{12}$ either in normal hierarchy 
with $P_H=1$ or in inverted hierarchy), except for the case of normal hierarchy with $P_{H}\simeq 0$, 
where $F_{\nu_\mu}+F_{\nu_\tau}\simeq F^0_{\nu_e}$ dominates. Therefore, we expect a signal variation 
by a factor $\sim \sin^2\theta_{12}\sigma_{\nu_e}/\sigma_{\nu_x}\sim 2$ (where $\sigma$ is the elastic 
cross section), due to our ignorance of the hierarchy and of $P_H$. Indeed, at 10~kpc we estimate 
from $\sim12$ to $\sim22$ events during the time interval of the neutronization peak ($t\in[42,47]$~ms 
in the SN simulation, see Fig.~\ref{lum}). The smallness of this time interval and the forward-peak 
signature of elastic events make this event sample basically background-free. In conclusion, a 0.4~Mton 
detector might observe the neutronization signal from a galactic supernova with a typical statistical 
significance of 3.5--4.7 standard deviations, assuming the emission model in Fig.~\ref{lum}. At the 
distance of the Large Magellanic Cloud, however, the sensitivity drops dramatically [$O(1)$ event, 
see Fig.~\ref{log}].

Let us now consider the (total) contributions from elastic scattering, absorption on oxygen, and 
inverse beta decay, integrated over the time interval $t\in[0,14]$~s. For elastic scattering, the 
slanted band in Fig.~\ref{log} is characterized by the smallest vertical spread, since the contributions 
from all flavors tend to partly cancel flavor transition effects. In a sense, the elastic scattering 
event sample ($\sim 4\times 10^3$ events at 10 kpc, easily separable through forward-direction cuts) 
could be used as a reference value to estimate the original, unoscillated neutrino flux. 
The cross section for $\nu_e$ and $\overline\nu_e$ absorption on oxygen becomes rapidly larger than 
the elastic scattering cross sections above $\sim 20$~MeV (see Fig.~\ref{xse}), where the adopted energy 
spectra are still sizeable (see Fig.~\ref{spe}). At $d=10$~kpc, we obtain about $1.1\times 10^4$ events 
($\pm20\%$ uncertainty from ignorance of the hierarchy and of $P_H$), with a significant contribution 
from $\nu_e$. If the case of normal hierarchy, the $\nu_e$ contribution might provide a handle to flavor 
transitions not observable in the ``canonical'' $\overline\nu_e$ channel (inverse beta decay), as we shall 
discuss in the context of shock-wave signatures. We remind that $e^\pm$ events from (anti)neutrino 
absorption on oxygen are slightly backward peaked \cite{Kolb}.

Finally, the inverse beta decay channel will provide, in general, such a large statistics in a 0.4~Mton 
Cherenkov detector ($\sim 2\times 10^{5}$ events at 10 kpc) that a handful of events might be seen even 
at a distance as large as $\sim 1$~Mpc. The high statistics available for a galactic supernova explosion 
will allow many possible spectral analysis, examples of which will be given in Sec.~IV in the context of 
shock-wave effects. Here we simply notice that, since the angular distribution of $e^+$ from inverse 
beta decay is slightly forward peaked (see, e.g., \cite{Stru}), a statistical separation from oxygen 
absorption events appears quite feasible, not only for the total event sample, but also for subsamples 
(e.g., time or energy bins). For instance, we have checked that, for an $O(10^5)$ inverse beta decay 
event sample, the slope of the positron distribution in $\cos\vartheta$ (where $\vartheta$ is the 
scattering angle between $\bar\nu_e$ and $e^+$) changes significantly---with respect to the very 
small statistical errors---when $O(10^4)$ oxygen events are added (not shown); further studies 
are needed, however, to make quantitative statements in this sense. 

\subsection{Supernova relic neutrinos}

In this section we study the flux of neutrinos coming from all past 
core-collapse supernovae (so-called 
Supernova Relic Neutrinos, SRN). Since the Super-Kamiokande SRN upper limit is only a factor 
of a few above typical flux estimates \cite{Male}, the SRN detection through inverse beta decay 
appears a feasible goal in Mton-class detectors, especially if the background can be suppressed 
by capturing the neutron with gadolinium \cite{GADZ}. SRN detection might be even closer
\cite{Stri} in view of recent experimental data from the Galaxy Evolution Explorer (GALEX)
mission \cite{GALE}, which suggest an increase of the 
estimated star formation rate (SFR) and thus of the SRN flux \cite{Stri}. In the following,
we present quantitative estimates of the SRN signal (for a 0.4 Mton detector), based on the emission 
model discussed in Sec.~II~A and on the same astrophysical input as we used in \cite{Deca}, except that 
we adopt here the lowest SFR function allowed by the (dust-corrected) GALEX and other data 
(i.e., the curve labeled as ``$\min\,A_\mathrm{FUV}$'' in Fig.~5 of \cite{GALE}), consistently 
with the ``concordance SFR'' advocated in \cite{Stri}. Our background estimates refer to the 
Kamioka site, for definiteness.

\begin{figure}[t]
\vspace*{-0.0cm}\hspace*{-0.2cm}
\includegraphics[scale=0.8]{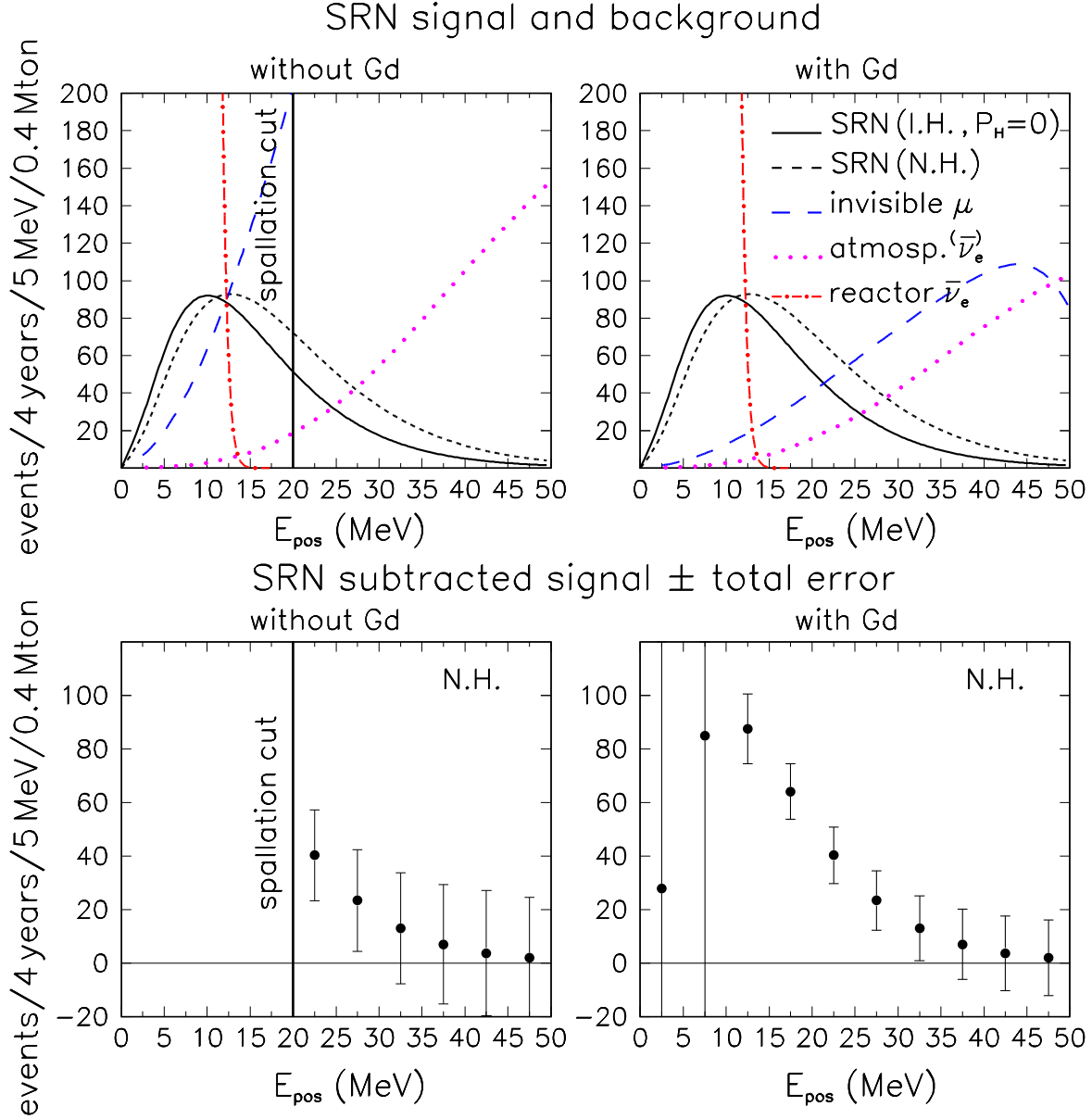}
\vspace*{+0.2cm} \caption{\label{rel}
\footnotesize\baselineskip=4mm Supernova relic neutrino signal and background
for a hypothetical 0.4 Mton detector at the Kamioka site, without and with
gadolinium (left and right panels, respectively). Upper panels: Energy spectra
of the signal and of atmospheric and reactor neutrino backgrounds.
Lower panel: subtracted signal with total statistical errors (for an exposure of 4 years).}
\end{figure}

Figure~\ref{rel} shows, in the upper left panel, the absolute relic neutrino spectrum 
expected in the two extreme cases of normal hierarchy and of inverted hierarchy with 
$P_H=0$, the latter being the most favorable in term of event rates. The abscissa is 
labelled by the measured positron energy. The vertical scale gives the number of events in 
5~MeV bin of the measured positron energy, for an exposure of (0.4~Mton)$\times$(4~years); 
for the sake of clarify, however, are shown as continuous curves rather than as 5-MeV bin 
histograms. In the upper left panel, we also show the background from invisible muons 
(with normalization taken from the SK measurement \cite{Male}, rescaled to 0.4~Mton), and 
the background from low-energy atmospheric $\nu_e$ and $\overline\nu_e$, which we have 
calculated using the recent FLUKA fluxes \cite{FLUK}. We remind that in the SRN energy range,
the $\overline\nu_e$'s interact dominantly through inverse beta decay and subdominantly through 
absorption on oxygen, while $\nu_e$'s dominantly scatter on oxygen. For simplicity,
we have assumed no atmospheric $\nu_e$ or $\overline\nu_e$ oscillations; however,
as also remarked in \cite{Deca}, we stress that oscillation effects induced by the
``solar'' squared mass difference $\delta m^2$ cannot be totally neglected in this 
context, since $\delta m^2 R_\oplus/E\sim O(1)$ in the energy range of Fig.~\ref{rel}. 
The reactor neutrino background has been estimated on the basis of the spectrum shape in 
\cite{Reac} (appropriate to parametrize the high-energy tail of the reactor spectrum), 
with total normalization provided by the observed flux in the KamLAND experiment 
\cite{KamL}. Finally, a $\sim20$ MeV hard cut to reject spallation events \cite{Male} 
is also shown. In the upper right panel (as compared to the upper left one) we assume 
that inverse beta decay events can be tagged by neutron capture in gadolinium.
In this case, although the reactor background cannot be reduced (having the same signature as 
SRN events), spallation events are fully rejected. Concerning atmospheric neutrinos, 
it is estimated that invisible muon events and $\nu_e$ oxygen absorption events can be 
reduced by a factor of $\sim 5$ with gadolinium \cite{Priv}.

From the comparison of the upper panels in Fig.~\ref{rel} we learn that, even with gadolinium, 
the extraction of the SRN signal requires a careful subtraction of (supposedly known) backgrounds 
from the total signal. Therefore, it is important to improve as much as possible the knowledge
of both the shape and the normalization of the background. In all cases, the shape is partly 
controlled by the energy resolution, which must be known very well; in fact, any increase or 
uncertainty in the resolution width $\Delta$ (due, e.g., to variation in the phototube coverage) 
will make some background event ``leak'' in the SRN sample, especially from the reactor neutrino 
sample, characterized by a steeply falling spectrum. Further studies of atmospheric neutrino fluxes 
\cite{FLUK} and of their interactions in oxygen \cite{Kolb} in the low-energy regime will also 
be beneficial to the reduce systematic uncertainties in the extraction of the SRN signal.

If we neglect systematics, the SRN number of event $N_S$ (derived from $N_\mathrm{tot}=N_S+N_B$, 
where $N_B$ is the estimated number of background events) is affected by an error 
$\pm\sqrt{N_\mathrm{tot}}$. Figure~\ref{rel} shows, in the lower panel, the prospective values of 
$N_S\pm\sqrt{N_\mathrm{tot}}$, for the less favorable case of normal hierarchy. The impact of 
gadolinium is evident both as a reduction of the error bars and as a reduction of the energy
threshold for SRN detectability. However, even without gadolinium, a $\sim 2$--$3\sigma$ SRN signal could 
emerge after an exposure of 4 years with 0.4 Mton. The situation would be more favorable in the case 
of inverted hierarchy with $P_H=0$ (not shown in the lower panels of Fig.~\ref{spe}).

In conclusion, there are good prospects to reveal a SRN signal with an exposure of a few years 
in a 0.4~Mton detector, especially (but not necessarily) with the addition of gadolinium. Further
studies are needed, however, for a better characterization of the background. In particular, 
in the SRN energy range, the tails of the reactor and atmospheric spectra, and of the energy
resolution function, need to be under control to avoid migration of events.

\section{Observables probing shock waves}

In this Section we study possible signatures of shock-wave effects that might be seen in a 0.4~Mton 
detector. We assume a galactic supernova ($d=10$ kpc) and consider first the shock signatures
in the absolute spectra of inverse beta decay events. We show then that a specific spectral ratio 
can actually monitor the time dependence of the crossing probability $P_H$, thus providing real-time 
information about the density profile. Finally, we show how events from elastic scattering and 
absorption on oxygen can further help the discrimination of shock-wave effects.

\subsection{Absolute time spectra from inverse beta decay}

Inverse beta decay events can be sensitive to shock-wave effects on $P_H$ only
in the case of inverted hierarchy [see Eqs.~(\ref{FnubarNH}) and (\ref{FnubarIH})]. 
However, inverted hierarchy is a necessary but not sufficient condition. 
For $\overline\nu_e$'s with $E\sim E_c$, where $E_c$ is the critical energy
introduced in Sec.~II~C ($E_c\simeq 20$ MeV for the our adopted emission model),
flavor transition effects will be largely cancelled, including matter effects
along the shock profile. No such cancellation is expected, however, at higher
energy (say, 45 MeV). Therefore, it makes sense to compare the expected
time spectra of events generated by neutrinos of relatively ``low'' energy
($E\sim E_c\sim 20$~MeV) and relatively ``high'' energy ($E\sim E_H\sim 45$~MeV)
for both normal and inverted hierarchy, with and without shock in the latter case.

\begin{figure}
\vspace*{-0.0cm}\hspace*{-0.2cm}
\includegraphics[scale=0.85]{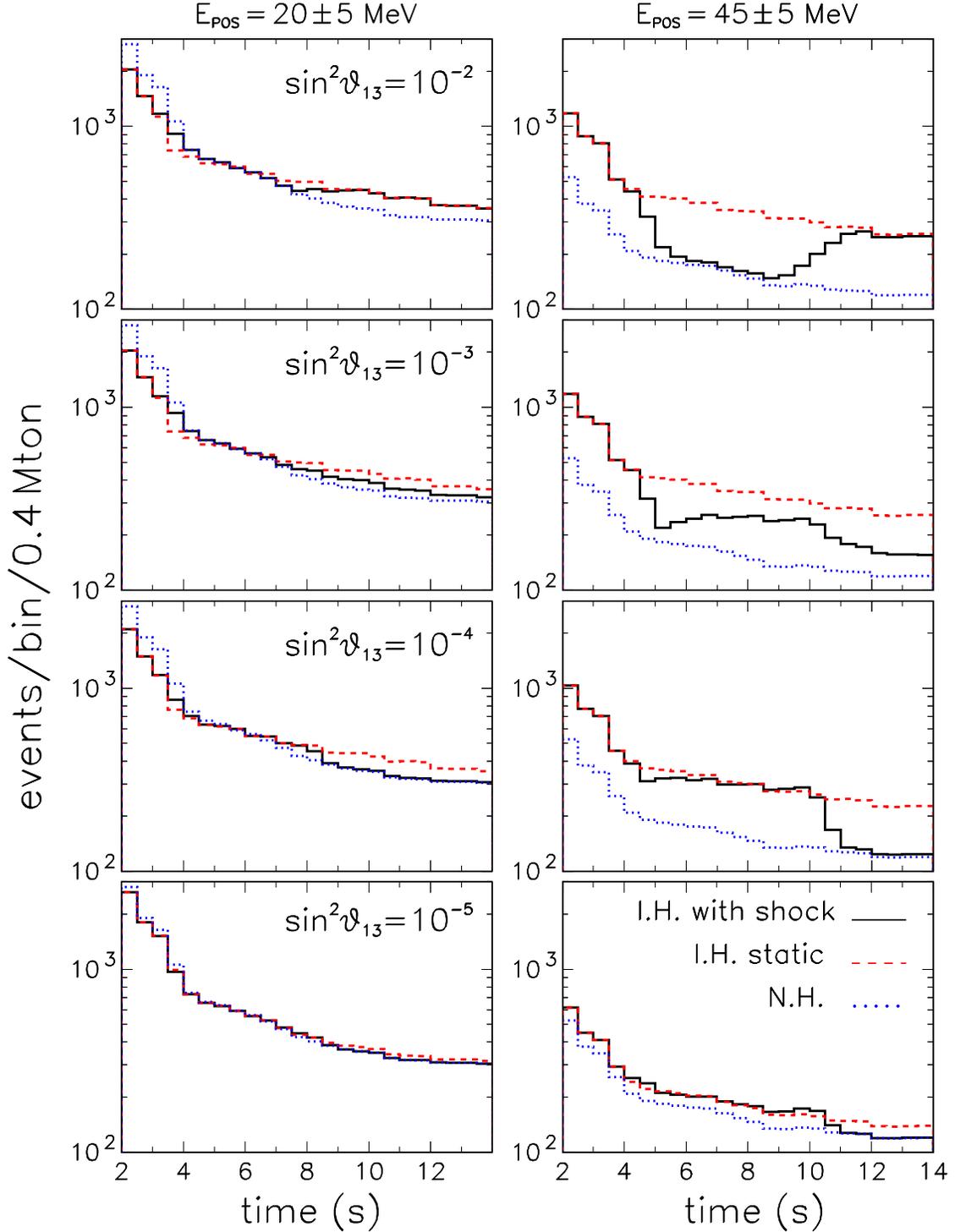}
\vspace*{-0.2cm} \caption{\label{fp1}
\footnotesize\baselineskip=4mm Absolute time spectra of events from inverse
beta decay in a 0.4 Mton detector, in the presence of forward shock only,
for four representative
values of $\sin^2\theta_{13}$. The solid, dashed, and dotted histograms
refer to calculations in inverted hierarchy with shock effects, inverted hierarchy
with a static density profile, and normal hierarchy, respectively. In each panel, 
the left (right) panels refer to the energy bin 
$E_\mathrm{pos}=20\pm5$ ($45\pm 5$) MeV.}
\end{figure}

\begin{figure}
\vspace*{-0.0cm}\hspace*{-0.2cm}
\includegraphics[scale=0.85]{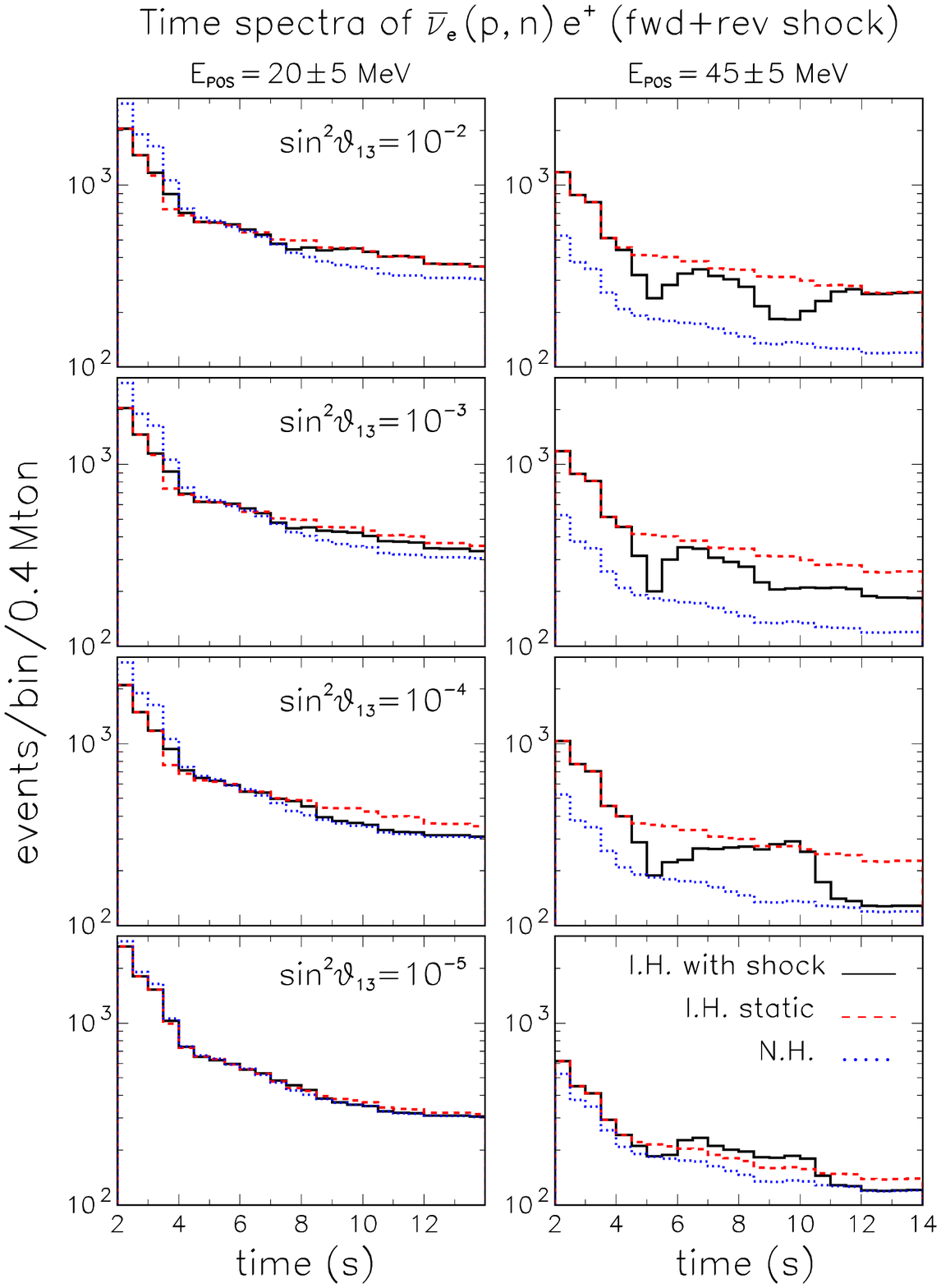}
\vspace*{-0.2cm} \caption{\label{fp2}
\footnotesize\baselineskip=4mm As in Fig.~\ref{fp1}, but for the case of forward
plus reverse shock.}
\end{figure}

Figures~\ref{fp1} and \ref{fp2} show absolute time spectra of events for the cases of forward 
shock only and of forward plus reverse shock, respectively, for the 
same representative values of $\sin^2\theta_{13}$ used in Fig.~\ref{pro}.
In both Fig.~\ref{fp1} and \ref{fp2}, the left (right) panels 
represent the time evolution of
the number of events within the positron energy bin $E_\mathrm{pos}=20\pm5$~MeV,
($E_\mathrm{pos}=45\pm5$~MeV)
for the cases of normal hierarchy, inverted hierarchy with static profile,
and inverted hierarchy with dynamical shock profile.  
The number of events in the bin
$E_\mathrm{pos}=45\pm5$~MeV is generally smaller than for $20\pm 5$~MeV. The time 
spectra in the bin $E_\mathrm{pos}=45\pm5$~MeV show strong signatures of shock-wave 
effects for inverted hierarchy, in the form of non-monotonic time variations of 
the event rate, especially at relatively high values of $\sin^2\theta_{13}$ and in 
the case of forward plus reverse shock. The amplitude of such variations is 
approximately contained within the two extreme cases of inverted hierarchy with 
no shock and of normal hierarchy. Conversely, the shock-induced time variations 
in the bin $E_\mathrm{pos}=20\pm5$ MeV are significantly smaller, and could be
hardly characterized a priori as non-monotonic. This feature can provide
a sort of ``postdiction'' of the critical energy $E_c$, should real shock-wave
effects be seen in a future supernova explosion: $E_c$ could be identified
as the energy where nonmonotonic time variations of the observed $\bar\nu_e$ signal 
are most suppressed (as compared with variations at higher energies). This empirical
definition of $E_c$ is sufficient for our purposes and, in any case, cannot
be made more precise, both because $E_c$ may fluctuate with time and because the 
detector energy resolution function has a nonnegligible width. Finally, we observe 
that the high number of events in each time bin of Figs.~\ref{fp1} and \ref{fp2} 
excludes that the 
spectral variations (or their absence) can be obscured by statistical fluctuations.

\subsection{Low-to-high energy ratio of time spectra}

The results discussed in the previous section suggest that the signatures of shock wave 
effects can be enhanced by comparing time spectra at the critical energy $E_c$, where, one
has by construction
\begin{equation}
\label{Ecritical} F_{\overline\nu_e}(E_c, t)\sim F^0_{\overline\nu_e}(E_c, t)\sim
F^0_{\nu_x}(E_c, t)
\end{equation}
at any $t$ [see Eqs.~(\ref{Ec}) and (\ref{Ec2})], with time spectra at significantly
higher energy $E_H$, where one expects $F^0_{\overline\nu_e}\ll F^0_{\nu_x}$
(see Figs.~\ref{lum} and \ref{spe}) and thus
\begin{equation}
\label{FNH}
F_{\overline\nu_e}(E_H, t)\simeq \sin^2\theta_{12}F^0_{\nu_x}(E_H, t)
\end{equation}
for normal hierarchy [Eq.~(\ref{FnubarNH})] and
\begin{equation}
\label{FIH}
F_{\overline\nu_e}(E_H, t)\simeq [1-\cos^2\theta_{12}P_H(E,t)] F^0_{\nu_x}(E_H, t)
\end{equation}
for inverted hierarchy [Eq.~(\ref{FnuNH})]. It is convenient to choose
$E_H$ so that the flux reduction from $E_c$ to $E_H$ 
[$F^0_{\nu_x}(E_H)/F^0_{\nu_x}(E_c)\ll 1$] is roughly compensated by the
increase of the inverse beta decay cross section [$\sigma(E_H)\gg \sigma(E_c)$],
\begin{equation}
\label{factor}
\frac{F^0_{\nu_x}(E_c)}{F^0_{\nu_x}(E_H)}\,\frac{\sigma(E_c)}{\sigma(E_H)}\sim 
1\ .
\end{equation}
With such choice for $E_H$, one can derive from the previous equations the following approximate
ratio of inverse beta decay events at $E_c$ and $E_H$: 
\begin{equation}
\label{NNH}
\frac{N(E_c, t)}{N(E_H, t)}\sim \frac{1}{\sin^2\theta_{12}}
\end{equation}
for normal hierarchy and
\begin{equation}
\label{NIH}
\frac{N(E_c, t)}{N(E_H, t)}\sim \frac{1}{1-\cos^2\theta_{12}P_H(E_H, t)}
\end{equation}
for inverted hierarchy.

For our adopted emission model, the condition in Eq.~(\ref{factor}) is fulfilled at 
$E_H\sim 45$~MeV, which explains a posteriori our choice for the ``high'' energy bin
in Figs.~\ref{fp1} and \ref{fp2}. Other choices for $E_H$ around 45~MeV would simply provide an overall 
factor on the right hand side of Eqs.~(\ref{NNH}) and (\ref{NIH}), with no substantial 
change in the following discussion.

\begin{figure}[t]
\vspace*{-0.0cm}\hspace*{-0.2cm}
\includegraphics[scale=0.92]{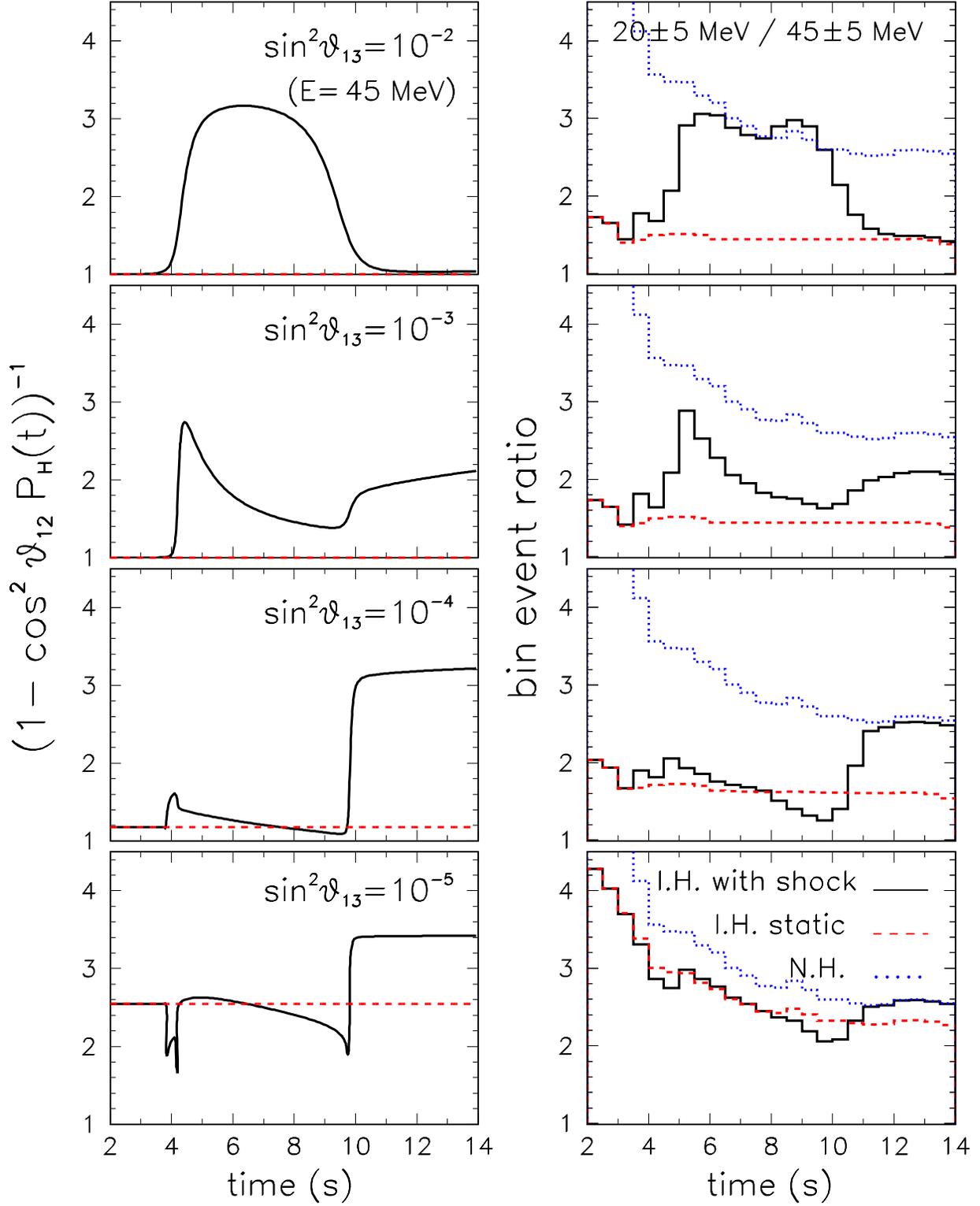}
\vspace*{-0.2cm} \caption{\label{fwd}
\footnotesize\baselineskip=4mm Time dependence of the ratio of 
events between the energy bins $E_\mathrm{pos}=20\pm 5$ MeV and
$E_\mathrm{pos}=45\pm 5$ MeV (right), compared with the function 
$[1-\cos^2\theta_{12}P_H(E_H,t)]^{-1}$ with $E_H=45$~MeV (right).  
It appears that, for inverted hierarchy, the event ratio on the right
tracks rather well the function structures on the left (up to smearing 
effects), thus providing a ``shock wave monitor'' in real time.}
\end{figure}

Figure~\ref{fwd} shows the ratio of events in the $20\pm 5$~MeV bin with respect to events 
in the $45\pm 5$~MeV bin, as compared with the function $[1-\cos^2\theta_{12}P_H(E_H, t)]^{-1}$. 
Notice that this function has the same qualitative behavior as $P_H(E_H, t)$ itself (see Fig.~\ref{pro}). 
It can be seen that, in inverted hierarchy, shock effects are faithfully monitored by the chosen ratio 
of events, i.e., there is a striking correspondence (even in absolute values, within a factor of 
2 or better) between the solid histograms on the right and the solid curves on the left in 
Fig.~\ref{fwd}, up to smearing effects due to the detector resolution. For inverted hierarchy 
with no shock (dashed curves and histograms) there is also a reasonable correspondence of the event 
ratio on the right with the constant value on the left. For normal hierarchy, as expected, any 
information on $P_H$ is lost. Therefore, in the case of inverted hierarchy, the ratio of events at 
``high energy'' and ``critical energy'' appears as a useful tool to track the main variations (and 
possibly the absolute value) of the crossing probability $P_H$, from which one could get precious 
information about the density gradient along the shock profile.

\begin{figure}
\vspace*{-0.0cm}\hspace*{-0.2cm}
\includegraphics[scale=0.92]{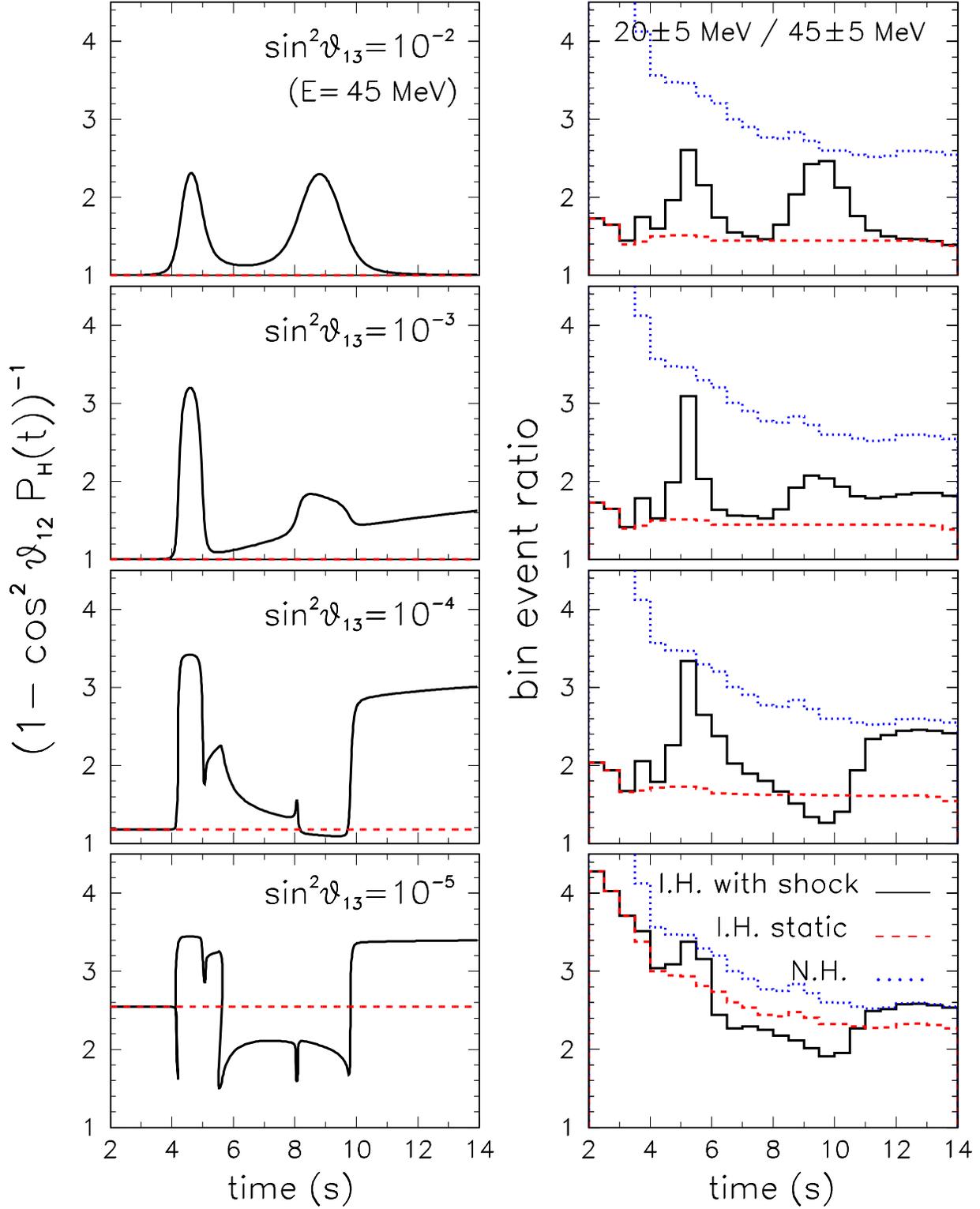}
\vspace*{-0.2cm} \caption{\label{rev}
\footnotesize\baselineskip=4mm As in Fig.~\ref{fwd}, but for the case of
forward+reverse shock.}
\end{figure}

Figure~\ref{rev} is analogous to Fig.~\ref{fwd}, but for the the case of forward plus reverse shock. 
Despite the more complex structure of the crossing probability function, its behavior is faithfully 
tracked by the ratio of events shown on the right (for inverted hierarchy), except perhaps at the 
lowest value of $\sin^2\theta_{13}$ shown. In any case, the main spikes and valleys of the function 
$[1-\cos^2\theta_{12}P_H(E_H, t)]^{-1}$ are clearly reproduced by the $(20\pm5)/(45\pm5)$~MeV event 
ratio, and cannot be mimicked by the very small statistical fluctuations (not shown). Of course, 
relations in Eqs.~(\ref{factor}), (\ref{NNH}), and (\ref{NIH}) are only approximated, and this 
is the reason why, for example, in the case of normal hierarchy (dotted line in Figs.~\ref{fwd} and 
\ref{rev}) or inverted hierarchy with static profile and $\sin^2\theta_{13}=10^{-5}$ (dashed line 
in the lower right panels of Figs.~\ref{fwd} and \ref{rev}) the ratio does not appear constant 
in time. 

Of course, when the next observable
supernova will explode, we will not know a priori the values of $E_c$ and 
of $E_H$ fulfilling Eqs.~(\ref{Ecritical}) and (\ref{factor}). However, if an imprint of shock waves 
is seen in the absolute spectra through nonmonotonic time variations, $E_c$ can be found by seeking 
the (low) energy range where such variations are suppressed, as already mentioned. Concerning $E_H$, 
its ignorance is not crucial to track the qualitative behavior of $P_H(t)$: histograms very similar 
to those shown in the right panels of Figs.~\ref{fwd} and \ref{rev} would be obtained by using, e.g., 
the $40\pm5$ MeV or $50\pm5$ MeV bin instead of $45\pm 5$~MeV, up to an overall rescaling of the y-axis 
(not shown).

\subsection{Time spectra from interactions on oxygen and on electrons}

\begin{figure}[t]
\vspace*{-0.0cm}\hspace*{-0.2cm}
\includegraphics[scale=0.67]{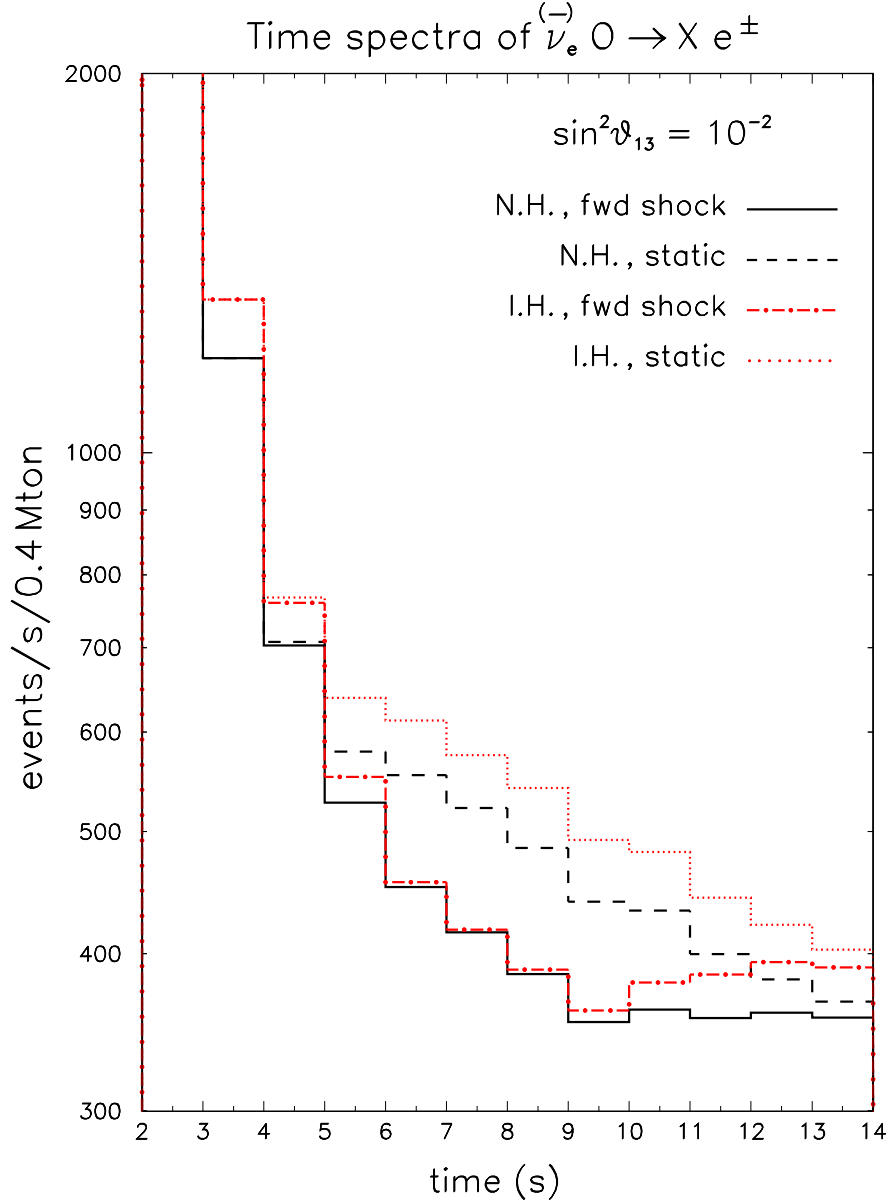}
\vspace*{-0.0cm} \caption{\label{oxy}
\footnotesize\baselineskip=4mm Time spectra of total oxygen absorption events for
normal and inverted hierarchy, and for static and shock profiles (with
$\sin^2\theta_{13}=10^{-2}$).}
\end{figure}

Both $\nu_e$ and $\overline\nu_e$ contribute to oxygen absorption events. Therefore, shock 
wave imprints are expected for both inverted and normal hierarchy. Figure~\ref{oxy} shows the 
time spectra of such events for both normal and inverted hierarchy, using either the static or 
the forward shock-wave profile for $\sin^2\theta_{13}=10^{-2}$. It can be seen that, in the 
presence of the shock, the event decrease in time is first steepened and then flattened for 
both hierarchies; in the case of inverted hierarchy, there is even a small dip at 9--10 seconds. 
Therefore, if the time decrease of the event rate is understood from simulations or from 
independent experimental information, the identification of a steepening+flattening behavior 
in oxygen events might be taken as a signal of shock passage (in the absence of alternative 
explanations).

\begin{figure}[t]
\vspace*{-0.0cm}\hspace*{-0.2cm}
\includegraphics[scale=0.67]{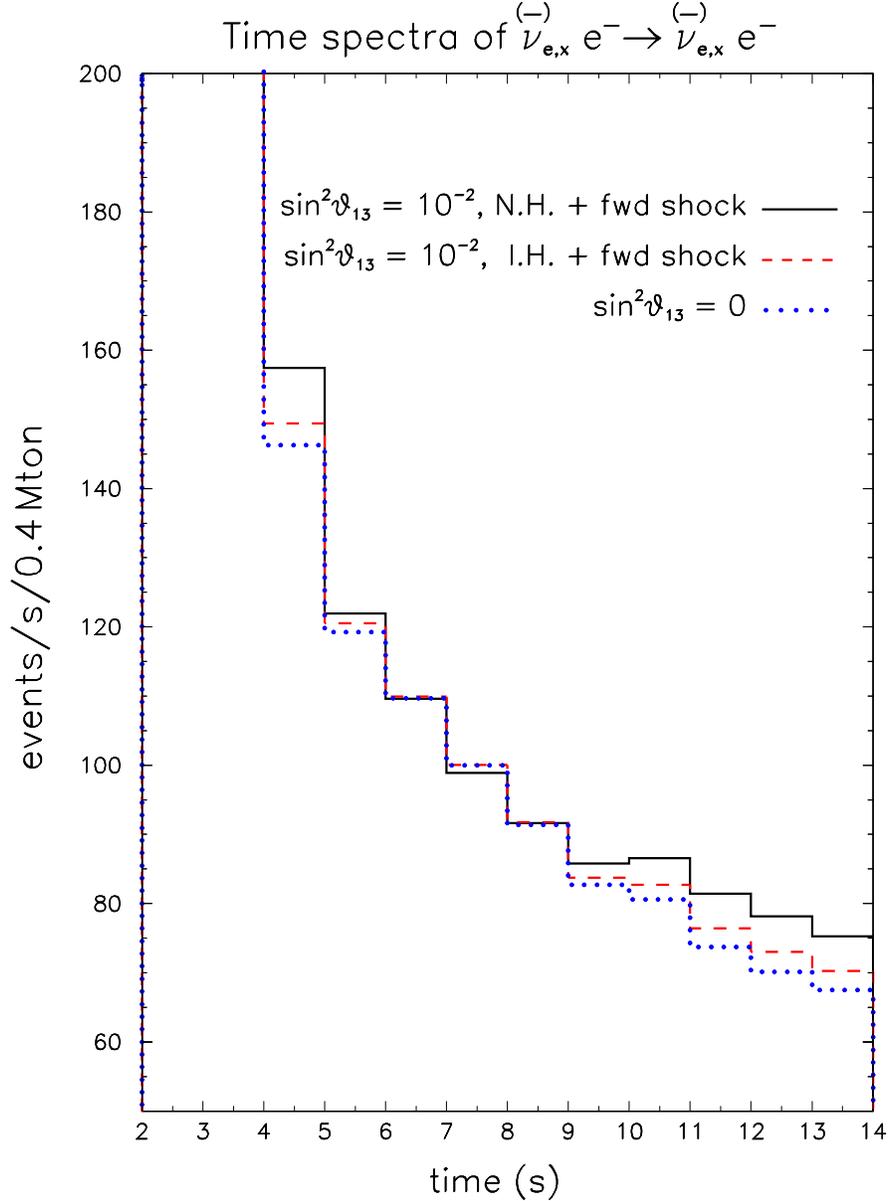}
\vspace*{+0.0cm} \caption{\label{ela}
\footnotesize\baselineskip=4mm Time spectra of total elastic scattering events
in normal and inverted hierarchy (with forward shock profile), 
for $\sin^2\theta_{13}=10^{-2}$ and 0.}
\end{figure}

Independent experimental information on the time decrease of the (anti)neutrino flux can be 
gained through elastic scattering events. Since neutrinos of all flavors contribute to this
sample, partial cancellation of oscillation effects occurs, as remarked in Sec.~III~A for the total 
rate in this class of events. Figure~\ref{ela} shows that the partial cancellation persists 
also in the time spectra, i.e., there are no large variations between normal and inverted
hierarchy and with or without crossing probability effects ($\sin^2\theta_{13}=10^{-2}$ or 0). 
Therefore, the elastic scattering sample is expected to track the overall decrease of flux in 
time, rather independently of flavor transition effects. This ``standard candle'' will be useful 
for comparison with the oxygen and inverse beta decay sample, where shock-wave effects can instead 
range from moderate to strong.

\section{Summary and Conclusions}

In this work we have analyzed the discovery potential of a 0.4~Mton water-Cherenkov 
detector in providing information on several time dependent and independent observables related 
to (extra)galactic core-collapse supernovae. We have considered elastic scattering 
on electrons as well as inelastic scattering on protons and on oxygen in the detector.

Concerning time independent observables, we have calculated the total number of events as a function 
of the distance of the exploding star for each detection channel during: a) the last pre-supernova 
(silicon burning) phase; b) the neutronization burst; and c) the core collapse  
phase. We have shown that the low energy ($E\lesssim 5$~MeV) thermal neutrinos emitted during the Si 
burning phase a couple of days before the explosion can be identified at $2\sigma$ level 
over typical backgrounds if the supernova is not too far ($d\lesssim 1$~kpc) and the detector
is loaded with gadolinium. The neutronization burst $\nu_e$'s can be identified with a significance 
$\gtrsim 3$ standard deviations if $d\lesssim 10$~kpc. From core collapse supernovae located near 
to the galactic center we expect up to $O(10^5)$ events from inverse beta decay events, and $O(10^4)$ 
events coming from elastic scattering and and oxygen absorption events. At the larger distance of,
e.g., the Andromeda galaxy (M31), we expect $O(10)$ events, thus opening a 
possible window for near extragalactic supernova surveys.

We have shown that with the addition of gadolinium, the energy range $[10,30]$~MeV can be fully 
exploited to detect the supernova relic neutrino flux at the level of several standard deviations
in a few years. However, even without gadolinium, 
the SRN background could emerge at $\sim 2$--$3\sigma$  level after an exposure of $\sim 4$ years, if 
the atmospheric background is under control.

We have then studied the time spectra of events in the range $t\in [2,14]$~s, in order to extract 
information on the neutrino crossing probability induced by the shock wave propagation in the stellar 
envelope during the cooling phase. In particular, we have considered the high statistic sample of 
inverse beta decay events. In this context, it is useful to introduce the concept of
 ``critical energy'' ($E_c\simeq 20$~MeV for our reference model), where oscillation effect 
 cancel to a large extent. In this way, the time 
dependence of the rate of the events in a bin around the critical energy is mainly affected 
by the overall decrease of the neutrino luminosity, and is almost independent from the 
time dependence of the probability. We have shown that the ratio between the number of events 
in a bin close to the critical energy ($E_c=20\pm 5$~MeV) and those in a bin centered at 
a significantly higher energy (e.g., $E_H=45\pm 5$~MeV) can faithfully track the 
neutrino crossing probability 
(if the mass hierarchy is inverted), thus providing a ``real-time'' movie of shock wave 
effects into the stellar envelope. In particular, the case of forward shock only \cite{Schi} 
and of forward+reverse shock \cite{Reve} leave significantly different signatures 
in the time domain, and can thus be discriminated.

We have completed our work by considering
the time dependence of the neutrino event rates from oxygen absorption and from elastic scattering. 
The latter are almost independent from shock and flavor transition effects, and 
can thus be used to monitor the overall decay of neutrino 
luminosity. The former (oxygen) event rate is instead sensitive to the shock wave 
also in the case of normal 
hierarchy (due to the $\nu_e$ contribution). If a ``steepening-flattening'' behavior is observed
in the oxygen event time spectrum, it could be interpreted as an imprint of the shock wave 
passage, in both cases of normal and inverted hierarchy.

In conclusion, Megaton-class detectors offer unprecedented opportunities to study supernova
and neutrino properties through high-statistics studies of energy and time spectra of 
supernova neutrino events from different interaction channels. Directions for further studies  
may include detector simulations of the statistical separation between such channels, 
more refined calculations of the oxygen absorption differential cross section,
improved numerical simulations of supernova explosion and shock-wave behavior, a better
understanding of the unoscillated neutrino spectra in time and energy (and of their
associated uncertainties), and an accurate characterization 
of the atmospheric and reactor background in supernova relic neutrino searches. 

\acknowledgments

We wish to thank J.\ Beacom and M.\ Vagins for very  useful discussions about the ``GADZOOKS!'' 
proposal for gadolinium loaded water-Cherenkov detectors; 
T.\ Montaruli and G.\ Battistoni for providing us 
the low energy fluxes of atmospheric neutrinos; P.\ Hernandez and
M.\ Mezzetto for discussions about
the Fr{\'e}jus project; and M.\ Nakahata and 
R.\ Tom{\`a}s for useful comments on the manuscript.
A.M.\ thanks C.\ Volpe for the kind invitation to the International Workshop 
on ``Exploring the Impact of New Neutrino Beams'' (Trento, Italy, 2004).  We also acknowledge fruitful 
discussions with several participants at the conference Neutrino 2004 (Paris, France, 2004) and 
at the Neutrino Oscillation Workshop (Otranto, Italy, 2004). This work is supported in part by 
the Italian ``Istituto Nazionale di Fisica Nucleare'' (INFN) and by the ``Ministero dell'Istruzione,  
Universit\`a e Ricerca'' (MIUR) through the ``Astroparticle Physics'' research project.


\end{document}